\documentclass[aps,nofootinbib,superscriptaddress, showpacs,preprintnumbers,  nofootinbibt,onecolumn]{revtex4-2}
\usepackage{booktabs}
\usepackage[utf8]{inputenc}
\usepackage[T1]{fontenc}
\usepackage{amsmath}
\usepackage{amssymb}
\usepackage{graphicx}
\usepackage{caption}
\usepackage{subcaption}
\usepackage[most]{tcolorbox}
\usepackage{mathtools}
\usepackage{epsfig}
\usepackage{multirow}
\usepackage{eurosym}
\usepackage{dcolumn}
\usepackage{bm}
\usepackage{enumerate}
\usepackage{float}
\usepackage{epstopdf}
\usepackage{amsmath}
\usepackage{bm}
\usepackage{amsfonts}
\usepackage{amssymb}
\usepackage{graphicx}
\usepackage{alphalph,mathtools}
\usepackage{etoolbox}
\usepackage{color}
\usepackage{booktabs}
\usepackage{footnote}
\usepackage{makecell,tabularx}
\usepackage[leftcaption]{sidecap}
\usepackage{amssymb}
\usepackage{graphicx}
\usepackage{booktabs,dcolumn,caption}
\usepackage{rotating}
\usepackage{amsthm}
\usepackage{tikz}
\usepackage{caption}
\usepackage{subcaption}
\usepackage[most]{tcolorbox}

\usepackage{hyperref}
\hypersetup{colorlinks,citecolor=blue}
\hypersetup{colorlinks=true,linkcolor=red,filecolor=green,    urlcolor=blue}

\def\be{\begin{equation}}
	\def\ee{\end{equation}}
\def\bea{\begin{eqnarray}}
	\def\eea{\end{eqnarray}}

\begin{document}
	
	\title{Black hole formation in gravitational collapse and their astrophysical implications}
	\author{Annu Jaiswal}
	\email{annujais3012@gmail.com}
	
	\author{Rajesh Kumar}
	\email{rkmath09@gmail.com}
	\author{Sudhir Kumar Srivastava}
	\email{sudhirpr66@rediffmail.com}
	\affiliation{Department of Mathematics and Statistics, D.D.U.Gorakhpur University, Gorakhpur, India.}
	
	\author{Megandhren Govender}
	\email{megandhreng@dut.ac.za}
	\affiliation{Department of Mathematics, Faculty of Applied Sciences, Durban University of Technology, Durban 4000, South Africa}

	\author{Shibesh Kumar Jas Pacif}
	\email{shibesh.math@gmail.com}
	\affiliation{Pacif Institute of Cosmology and Selfology (PICS), Sagara, Sambalpur 768224, Odisha, India}

		\begin{abstract}
		In this work, we have investigated a novel aspect of black hole (BH) formation during the collapse of a self-gravitating configuration. The exact solution of the Einstein field equations is obtained in a model-independent way by considering a parametrization of the expansion scalar ($\Theta$) in the background of spherically symmetric space-time geometry governed by the FLRW metric. Smooth matching of the interior solution with the Schwarzschild exterior metric across the boundary hypersurface of the star, together with the condition that the mass function $m(t,r)$ is equal to Schwarzschild mass $M$, is used to obtain all the physical and geometrical parameters in terms of the stellar mass. The four known massive stars namely $R136a3$, $Melnick$, $R136c$, and $R136b$ with their known astrophysical data (mass, radius, and present age)  are used to study the physics of the model both numerically and graphically. We demonstrate that the formation of the apparent horizon occurs earlier than the singular state that is,  the model of massive stars would inevitably lead to the formation of a BH as their end state. We have conducted an analysis indicating that the lifespans of massive stars are closely related to their respective masses. Our findings demonstrate that more massive stars exhibit considerably shorter lifespans in comparison to their lighter counterparts. Thus, the presented model corresponds to the evolutionary stages of astrophysical stellar objects and theoretically predicts their possible lifespan. We have also shown that our model satisfies the energy conditions and stability requirements via Herrera's cracking method.

	\end{abstract}

	\maketitle
	\tableofcontents

	\section{Introduction}\label{sec1}
		The formation of black holes through gravitational collapse is a fascinating and important phenomenon in astrophysics. Black holes are regions in space-time where the gravitational pull is so strong that nothing, not even light, can escape from them. Black holes, particularly the end-states of gravitational collapse, have received considerable attention in recent years and have witnessed rapid theoretical developments as well as numerous astrophysical applications.  The eventual consequence of gravitational collapse in general relativity is a subject of immense significance and interest from the perspective of black hole physics \cite{c3}. The eventual outcome of continuous gravitational collapse culminating in either a black hole or naked singularity is intrinsically dependent on the nature of the initial data from which the collapse ensues \cite{c2}-\cite{c4}. Some recent works also revealed that there is a continuous collapse without any state of equilibrium known as "Eternal gravitational collapse" \cite{ne34a}-\cite{AJ223}.
	\par
From the singularity theorems of Hawking and Ellis through to the theoretical investigations of black holes by Bekenstein, Geroch, Joshi, and Penrose, amongst others, the end-states of continued gravitational collapse were merely mathematical excursions into general relativity \cite{misner,shapiro}. This changed in 2019 with the first photograph of a black hole shadow. This was a game changer in black hole physics in the sense that a theoretical/mathematical construct of general relativity revealed itself in physical reality. The LIGO-Virgo collaboration documented a gravitational event referred to as GW190814 which perplexed many researchers and continues to do so \cite{waves1,waves2,waves3}. The signal emanating from this event is thought to be the result of a compact binary coalescence of a $23.2^{+1.1}_{-1.0}{M_\odot}$ BH and a secondary component whose mass ranges from 2.50 to 2.67$M_{\odot}$. This is now believed to be the most unequal mass ratio to date, $0.112_{-0.009}^{+0.008}$ for a binary merger leading researchers to the conclusion that the secondary component is the lightest black hole or the heaviest neutron star to be observed in a binary system. The challenge for researchers is to come up with a salient model of a compact object with a mass exceeding 2.50 $M_{\odot}$ \cite{burgio}. Black hole physics in light of observational data of black shadows, gravitational waves, and supermassive black holes has led to the rethinking of the possible equation of states, the role of anisotropy and charge as well as the effects of rotation.  Black holes are characterized by an event horizon, which is a boundary beyond which nothing can return, and inside it, the gravitational forces are so strong that the fabric of space-time is severely curved, leading to the formation of a singularity at the center—a point of infinite density and curvature where laws of physics break down.
	\par
	The Oppenheimer-Snyder(OS) spherically symmetric collapse solution \cite{c5}, in which a dust cloud continues to collapse until it forms a black hole, is a well-known model that has served as the fundamental paradigm in black hole physics for simulating such a physical process.  OS initiated the study of homogeneous gravitational collapse with an FLRW-like metric and their model serves as the framework for the notion of the inevitable development of BH as far as solutions are concerned.  
	The OS collapse scenario was a highly idealized study. Herrera and co-workers \cite{herrera1},\cite{herrera2} have over the past three decades investigated the influence of pressure and dissipation in collapsing spheres. The most general collapse model in their framework consisted of a spherically symmetric, self-gravitating sphere in which the stellar interior was endowed with heat flow, shear, charge, and pressure. In the literature, the phenomenon of gravitational collapse and space-time singularities have been studied by a number of researchers in various approaches in general relativity and modified gravitational theories (\cite{c6}-\cite{AJ22} amongst other). 
 \par
	In general relativity, the motion of collapsing fluids is governed by a number of parameters, including shear tensor, vorticity tensor (which vanishes in the present case), acceleration vector, and expansion scalar ($\Theta$). The dynamics of the stellar system are determined by these kinematical quantities that evolve throughout the gravitational collapse. Recently, authors \cite{AJ223},\cite{RJ22},\cite{AJ22} have studied a new class of gravitational collapse with uniform expansion scalar, which may describe the interesting scenario of collapsing stellar systems and may also have many astrophysical consequences. The current work aims to examine the homogeneous gravitational collapse of perfect fluid distributions from entirely new approaches and discuss the solution of EFEs by using boundary conditions and astrophysical data. Recently, the authors have introduced some parameterization of expansion scalar $\Theta$ and studied the gravitational collapse of massive stars and their possible end-states\cite{AJ223}. In this work. we have introduced a novel $\Theta$-parameterization as a function of the scalar factor. We have considered the astrophysical stellar data of some known massive stars namely, R136b, R136c, Melnick 34A, and  R136a3, and then obtained solutions are discussed both analytically and graphically for these data points. Further, a comprehensive discussion of an apparent horizon, BH formation, and life -span of stars is carried out for the aforementioned stars. 
	The physical validity of the derived model is subjected to causality and stability conditions together with the energy conditions. 
	\par
	The paper is organized as follows: following the Introduction, we present the governing Einstein field equations (EFEs) for FLRW space-time metric with perfect fluid distributions in section II. The exterior region of the system is considered to be a vacuum described by the Schwarzschild solution. In section III we introduce a parameterization of the expansion scalar to obtain exact solutions of EFEs. In section IV, we discuss the dynamics of the collapsing model and provide estimates for the numerical values of model parameters from known stellar data. We further provide graphical analyses of the gross features of our model.  In Section V, we discuss the formation of the apparent horizon, black hole formation, and life-span of aforementioned stars. In Section VI, we discuss the energy conditions for the model together with their graphical representations. The stability analysis via the Herrera approach for our model is described in Section VII. We conclude with a discussion of our findings in section VIII.

	\section{General formalism of Collapsing Spherical Star with perfect fluid distribution}\label{sec2}
	\subsection{The metric and basic equations}\label{subsec1.1}
	We consider the spherically symmetric gravitational collapse of a stellar system (e.g., star) with interior geometry described by the FLRW metric
	\begin{equation}	
		ds_{-}^2 = h_{\alpha \beta}dx^{\alpha}dx^{\beta}+ R(t,r)^{2}d\Omega^2
		\label{eq1}
	\end{equation}	
	with $h_{\alpha \beta} = diag(-1, a^{2})$, $\alpha, \beta = 0, 1 $ and $x^{0} = t$, $x^{1} = r$ and $d\Omega^2 \equiv d\theta^2 + \sin^2\theta d\phi^2$ is the metric on the 2-sphere. Here $a(t)>0$ is the scale factor and $R(t,r)= r a(t)$ is the geometric radius of the collapsing system. The collapse process of the star is not only influenced by the space-time around it but also by the geometric component of its internal space-time geometry, which is governed by the metric (\ref{eq1}) and matter distribution inside the star. We consider the interior matter distribution as a perfect fluid which is described by the  following energy-momentum tensor  
	\begin{equation}
		T_{ij} = (p+\rho) V_j V_j + p g_{ij}
		\label{eq2}
	\end{equation}
	where $\rho$ and $p$ are the energy density and isotropic pressure respectively, and $V^i$ is four-vector velocity in co-moving coordinates $x^i = (t,r,\theta, \phi)$ satisfying $V^{i} V_{i} =-1$, where $i, j = 0,1,2,3$
	\par
	
	In the collapse scenario of a stellar system,  the motion of the fluid distribution is towards the core (center) of the star, therefore one generally assumes that $\dot{R} < 0$. The collapse rate of the fluid distribution (inside the star) is described by the expansion scalar
	\begin{equation}
		\Theta = V^i_{;i} = 3\frac{\dot{R}}{R} (<0)
		\label{eq3}
	\end{equation}
	where semicolon $(;)$ represents the covariant derivative and  dot $(.)$ denotes the derivative with respect to time $t$. 
	\par
	The Einstein field equation	
	\begin{equation*}
		R_{ij} - \frac{1}{2}  \mathcal{R} g_{ij}=\mathcal{X} T_{ij}
		\label{eq4}
	\end{equation*}
	for the considered system yields the following (where  $\mathcal{X} = \frac{8 \pi G}{c^2}$ )
	\begin{equation}
		\mathcal{X}\rho = \frac{1}{3} \Theta^2 
		\label{eq5}
	\end{equation}
	
	\begin{equation}
		\mathcal{X} p + \frac{1}{3} \Theta^2 + \frac{2}{3}\dot{\Theta} = 0 
		\label{eq6}
	\end{equation}
	\begin{equation}
		\mathcal{X} p + \left(\frac{1}{3} \Theta^2 + \frac{2}{3}\dot{\Theta} \right) \sin^2\theta = 0 
		\label{eq7}
	\end{equation}
	The mass-function $m(t,r)$ of the collapsing stellar system at any instant $(t,r)$ is given by \cite{me70} \\
	\begin{equation}
		m(t,r) = \frac{1}{2} R \left(1 + R_{,i} R_{,j} g^{ij}\right) = \frac{1}{18} \Theta^2 R^3 
		\label{eq8}
	\end{equation}	
	Here comma (,) denotes partial differentiation.	Also, in view of eqs.(\ref{eq5})-(\ref{eq6}) and (\ref{eq8}) we obtain
	\begin{equation}
		\dot{m} = -\frac{1}{2} \mathcal{X} p \dot{R} R^2
		\label{eq9}
	\end{equation}
	\begin{equation}
		m' = \frac{1}{2} \mathcal{X} \rho R' R^2
		\label{eq10}
	\end{equation}
	\par
	In the study of space-time manifolds, it is important to know if the geometry associated with these manifolds is smooth or not. By a smooth space-time manifold, we simply mean that it must have regular curvature invariants that are finite at all space-time points, or it must include curvature singularities, at least one of which is infinite. In many cases, one of the most useful ways to know this is by checking for the finiteness of the Kretschmann scalar curvature (KS) which sometimes is called the Riemann tensor squared, \cite{D92, SO02}
	\begin{equation}
		\mathcal{K} = R_{\alpha \beta \gamma \delta} R^{\alpha \beta \gamma \delta}
		\label{eq11}
	\end{equation}
	For the space-time manifold (\ref{eq1}), it gives
	\begin{equation}
		\mathcal{K} = \dfrac{12 \left(\dot{a}^4 + a^2 \ddot{a}^2\right)}{a^4}
		\label{eq12}
	\end{equation}    
	
	
	\subsection{Exterior metric and the boundary condition}\label{subsec1.2}
	In general relativity, the Schwarzschild metric is the unique spherically symmetric solution of the vacuum Einstein field equations. In that case, a spherically symmetric gravitational field in a vacuum exterior to a spherical body must be static and asymptotically flat (Birkhoff's theorem). Since the theorem's applicability is local and therefore it can be used as boundary conditions for any stellar system. In the present study, we consider the geometry of the exterior region of a star to be described by the vacuum Schwarzschild metric which can be cast as
	
	\begin{equation}
		ds^{2}_{+} = -\left(1-\frac{2M}{\xi}\right) dT^{2} + \frac{d\xi^{2}}{\left(1-\frac{2M}{\xi}\right)} + \xi^{2} d\Omega^{2}
		\label{eq13}
	\end{equation}
	where the exterior coordinates  $x_{+}^{i} =(T, \xi, \theta, \phi)$ and $M$ is the stellar mass (Newtonian mass) called the Schwarzschild mass.
	\par
	The boundary hypersurface $\Sigma$$(r = r_{0})$ divides the spherically symmetric stellar system into  the interior $(ds^{2}_{-})$ and exterior $(ds^{2}_{+})$ space-time regions. The junction conditions (boundary conditions) are essentially the continuity of the first and second fundamental forms (the intrinsic metric and extrinsic curvature) across the hypersurface $\Sigma$ and have been studied by Santos \cite{NO16}.  Moreover, the continuity of the first and second fundamental forms on $\Sigma$ yield the boundary conditions- (i) pressure vanishes at $\Sigma$, and (ii) the mass function m(t,r) must be equal to the Schwarzschild mass $M$ at $\Sigma$ i.e., 
	\begin{equation}
		m\left(t,r\right) \overset{\Sigma}{=} M
		\label{eq14}
	\end{equation}
	Theoretically, the boundary conditions are used to estimate the numerical value of arbitrary constants. Therefore, we can determine the arbitrary constants in terms of the mass of the star ($M$) by using the aforementioned condition (\ref{eq14}).
	
	\section{Parametrization of Expansion scalar}\label{sec3}
 
	The system of differential equations (\ref{eq5})-(\ref{eq7}) possess only two independent equations with three unknowns, viz.,  $a(t)$, $p(t)$, and $\rho(t)$. Therefore,  it requires one more constraint for the complete determination of the solution of the EFEs.  As it is well known, the EFEs do not completely determine the system as there is an extra degree of freedom that is usually closed by an Equation of state (parametrization of $p$ and $\rho$). In fact, a critical analysis of the solution in general relativity (or, in modified gravity theories) is the parametrization of physical/kinematical/geometrical parameters used by theoretical physicists \cite{glass79}-\cite{herrera2009}, \cite{AJ223},\cite{AJ22} amongst others. In the present study,  we have introduced a parametrization scheme of the kinematical parameter, i.e., the expansion scalar (\ref{eq15}). The motivation for such kinds of parametrization is as follows:
	\par
\begin{figure}[h]
		\centering
		\includegraphics[scale=0.90]{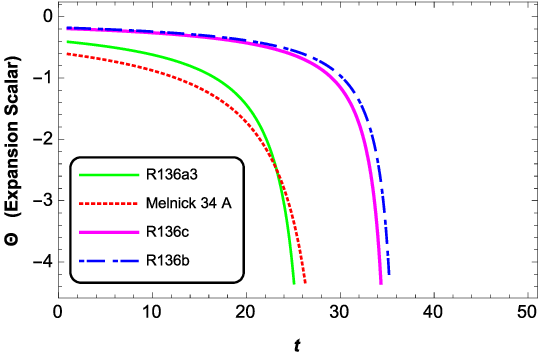}
		\caption{Collapsing configuration: Variation of the expansion scalar $\Theta$ versus time coordinate $t$ for the four massive stars corresponding to model parameter $\beta$ given in Table \ref{table1}.}
		\label{figA}
	\end{figure}

	In the evolution of the stellar system (e.g., star) due to nuclear fusion in the core of a star, it loses its equilibrium stage and starts to collapse under its own gravity. During the collapse process, the internal thermal pressure (which is due to nuclear fusion of hydrogen or helium ) decreases, and then the external gravitational pressure (which is due to the mass of the star) dominates over it. At the same time, the collapse rate of the star increases which tends to draw matter inward towards the center of the stellar configuration. According to GR, in the collapsing system, two kinds of motion (velocities) occur namely $\frac{\dot{R}}{R}$ which measures the variation of areal radius $R$ per unit proper time and, another $(\delta l)^{'}$, the variation of the infinitesimal proper radial distance, $(\delta l)$  between two neighboring fluid particles per unit proper time \cite{herrera2009}. The expansion scalar $\Theta$ is defined as the rate of change of elementary fluid particles which describes the collapsing rate of the fluid distribution in a stellar system. Critical analyses show that for collapsing configurations, $\Theta$ increases with $t$ (see figure \ref{figA}). Also, the FLRW homogeneous gravitational collapse requires the motion of fluid particles to be uniform, independent of $r$, and scale factor $a(t)$ decreases with $t$ as collapse proceeds (see figure \ref{figB}). Consequently, the expansion scalar increases as $a(t)$ decreases in a collapsing scenario.  It follows that a natural parameterization of $\Theta$, one can consider it as a function of $a(t)$, which precisely explains its notion and also depicts the collapsing configuration. Recently, we have studied various $\Theta$-parametrizations to solve the field equations that describe the collapse of stellar systems\cite{AJ223},\cite{AJ22}.\\ 
	In our present study, we introduce a novel parameterization of $\Theta$ as\footnote{In the collapsing process $\Theta <0$, and here the negative sign convention represents the motion of collapsing fluids towards the center of the star.}\\
\begin{equation}
		\Theta = -{a}^{-\beta}
		\label{eq15}
\end{equation} 	
	where $\beta$ is the model parameter to be determined for known astrophysical data. 
	
	\section{Dynamics of collapsing massive star}\label{sec4}
	\subsection{Exact solution of Einstein field equation}\label{subsec4.1}
	We consider the parametrization (\ref{eq15}) as an additional constraint to obtain the solution of field equations (\ref{eq5})-(\ref{eq7}). In view of eqs.(\ref{eq3}) and (\ref{eq15}) we have
	\begin{equation}
		\frac{3 \dot{a}}{a}+a^{-\beta }=0
		\label{eq16}
	\end{equation}
	and on solving differential equation (\ref{eq16}), it gives
	
	\begin{equation}
		a(t) = \left[\beta  \left(-\frac{  t}{3}+{k}\right)\right]^{\frac{1}{\beta }}
		\label{eq17}
	\end{equation}
	where $k$ is an integration constant. In order to determine the value of  $k$, we use the boundary condition (\ref{eq14}) on $\Sigma$ at $r=r_0$.
	
	\par
	 Assuming that the star begins to collapse at a moment $(t,r) = ( t_{0}, r_{0})$ which defines the initial boundary condition for the system. Then by using eqs.(\ref{eq3}), (\ref{eq8}) and (\ref{eq17}) into eq.(\ref{eq14}), we obtain
	\begin{equation}
		\frac{1}{18}  r_{0}^3 \left(k \beta-\frac{ t_{0}\beta}{3}\right)^{-2+\frac{3}{\beta }} = M
		\label{eq18}
	\end{equation}
	Solving (\ref{eq18}) for  $k$,  yields 
	
	\begin{widetext}
		\begin{equation}
			{k} =  \frac{2^{-\frac{\beta }{2 \beta -3 }} 3^{-1-\frac{2 \beta }{2 \beta -3}} \left(\frac{M}{r_{0}^3}\right)^{-\frac{\beta }{2 \beta -3}} \left(3+  2^{\frac{\beta }{2 \beta -3 }}  3^{\frac{2 \beta }{2 \beta -3 }}   \left(\frac{M}{r_{0}^3}\right)^{\frac{\beta }{2 \beta -3}} t_{0}\beta\right)}{\beta }
			\label{eq19}		
		\end{equation}
	\end{widetext}

	After substituting this value of $k$ into  eq.(\ref{eq17}), we obtain the scale factor
	
	\begin{equation}
		a(t)  = \left[18^{\frac{\beta}{3-2 \beta }} \left(\frac{M}{ r_{0}^3}\right)^{\frac{\beta }{3-2 \beta }}-\frac{t \beta  }{3}+\frac{ t_{0} \beta}{3}\right]^{\frac{1}{\beta }}
		\label{eq20}
	\end{equation}
	
	\begin{figure}[h]
		\centering
		\includegraphics[scale=0.97]{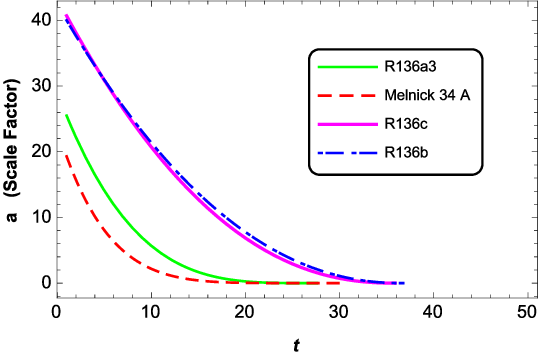}
		\caption{Variation of the scalar factor $a(t)$ versus time coordinate $t$ for the four massive stars, assuming initial coordinate $(t_0, r_0) = (1, 1)$.}
		\label{figB}
	\end{figure}

	Thus, in view of eq.(\ref{eq20}), we obtain the physical parameters from eq.(\ref{eq5})-(\ref{eq6})
	
	\begin{equation}
		\mathcal{X} \rho = \frac{1}{3 \left[18^{\frac{\beta}{3-2\beta}}\left(\frac{M}{r_{0}^3 }\right)^{\frac{\beta}{3-2\beta}}-\frac{t\beta  }{3}+\frac{ t_{0}  \beta}{3}\right]^{2}}
		\label{eq21}
	\end{equation}
	\begin{figure}
		\begin{subfigure}[a]{0.4\textwidth}
			{\includegraphics[width=7cm]{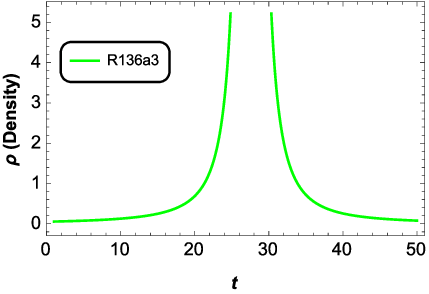}}
			\label{fig1}
		\end{subfigure}
		\begin{subfigure}[a]{0.4\textwidth}
			{\includegraphics[width=7.2cm]{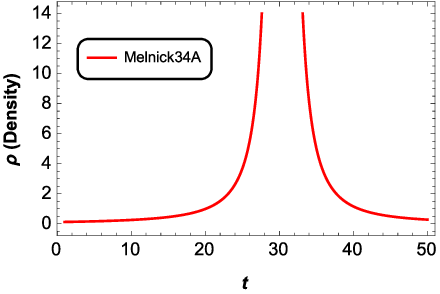}}
			\label{fig2}
		\end{subfigure}
		\begin{subfigure}[c]{0.4\textwidth}
			{\includegraphics[width=7cm]{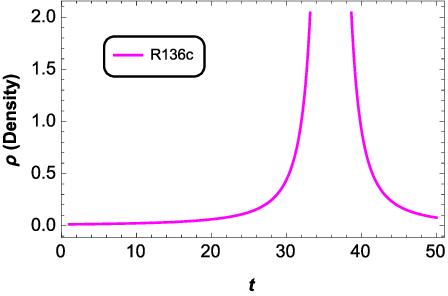}}
			\label{fig3}
		\end{subfigure}
		\begin{subfigure}[c]{0.4\textwidth}
			{\includegraphics[width=7cm]{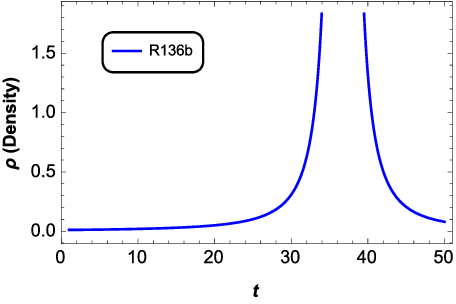}}
			\label{fig4}
		\end{subfigure}
		\caption {Variation of the density $(\rho)$ versus time coordinate $t$ for the four massive stars .}
		\label{figC}
	\end{figure}
	
	\begin{equation}
		\mathcal{X} p =\frac{2 \beta-3}{ \left[2^{\frac{\beta }{3-2 \beta }} 27^{\frac{1}{3-2 \beta }} \left(\frac{M}{ r_{0}^3}\right)^{\frac{\beta }{3-2 \beta }}- t   \beta  + t_{0}  \beta  \right]^2}
		\label{eq22}
	\end{equation}
	\begin{figure}
		\begin{subfigure}[a]{0.4\textwidth}
			{\includegraphics[width=7cm]{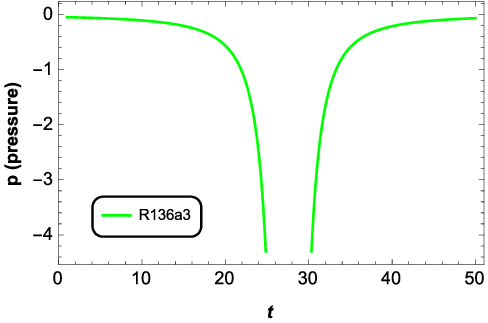}}
			\label{fig5}
		\end{subfigure}
		\begin{subfigure}[a]{0.4\textwidth}
			{\includegraphics[width=7.2cm]{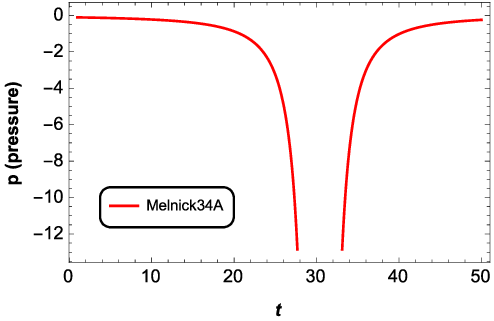}}
			\label{fig6}
		\end{subfigure}
		\begin{subfigure}[c]{0.4\textwidth}
			{\includegraphics[width=7cm]{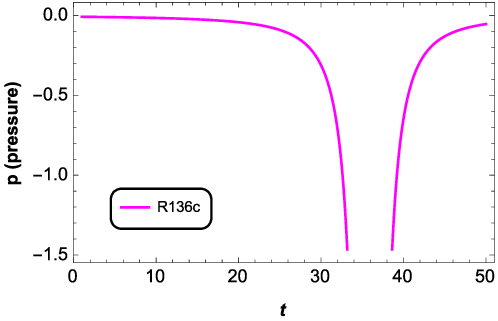}}
			\label{fig7}
		\end{subfigure}
		\begin{subfigure}[c]{0.4\textwidth}
			{\includegraphics[width=7cm]{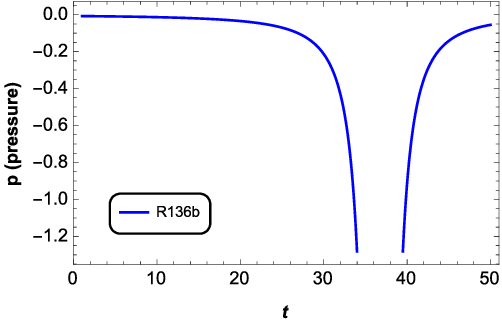}}
			\label{fig8}
		\end{subfigure}
		\caption{Variation of the pressure $(p)$ versus time coordinate $t$ for the four massive stars .}
		\label{figD}
	\end{figure}
	Also from eqs.(\ref{eq8})-(\ref{eq10}), the collapsing mass ($m$) and its variation with $t$ and $r$ are
	\begin{equation}
		m =\frac{1}{18} r^3 \left[18^{\frac{\beta }{3-2 \beta }} \left(\frac{M} {r_{0}^3}\right)^{\frac{\beta }{3-2 \beta }}-\frac{ t\beta}{3}+\frac{  t_{0}\beta}{3}\right]^{-2+\frac{3}{\beta }}
		\label{eq23}
	\end{equation}
	\begin{figure}
		{\includegraphics[scale=1.1]{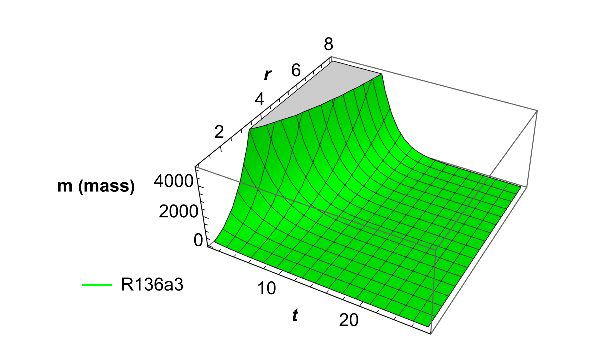}}
		\caption{Variation of the mass $(m)$ versus time coordinate $t$  and radial coordinate $r$ for the R136a3.}
		\label{figE}
	\end{figure}
	\begin{equation}
		\dot{m}=-\frac{1}{54}   r^3 \left(-2+\frac{3}{\beta }\right) \beta  \left[18^{\frac{\beta }{3-2 \beta }} \left(\frac{M}{r_{0}^3}\right)^{\frac{\beta }{3-2 \beta }}-\frac{t \beta  }{3}+\frac{t_{0}\beta  }{3}\right]^{-3+\frac{3}{\beta }}
		\label{eq24}
	\end{equation}
	\begin{figure}
		{\includegraphics[scale=1.1]{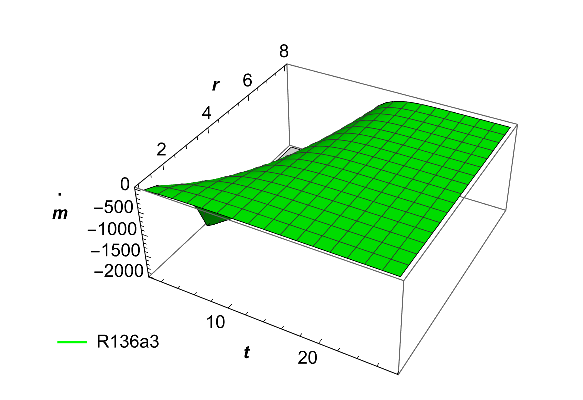}}
		\caption{Variation of rate of mass $(\dot{m})$ versus time coordinate $t$  and radial coordinate $r$ for the R136a3 .}
		\label{figF}
	\end{figure}
		\begin{equation}
			m'= \frac{1}{6} r^2 \left[18^{\frac{\beta}{3 - 2 \beta}} \left(\frac{M}{r_{0}^{3}}\right)^{\frac{\beta}{3-2\beta}}- \frac{t\beta  }{3} + \frac{t_{0}\beta }{3}\right]^{-2 + \frac{3}{\beta} }
			\label{eq25}
	\end{equation}
	\begin{figure}
		{\includegraphics[scale=1.1]{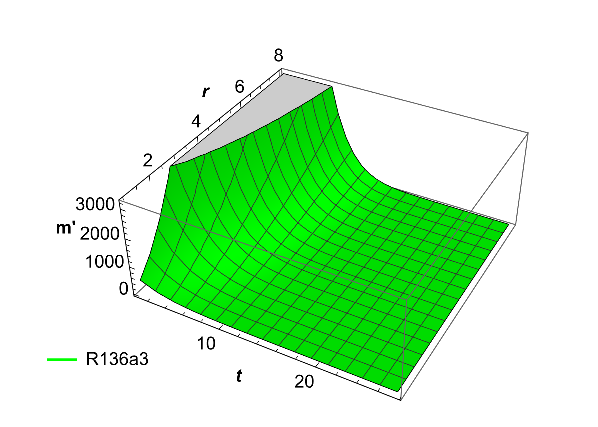}}
		\caption{Variation of the gradient of mass $(m')$ versus time coordinate $t$  and radial coordinate $r$ for the R136a3 .}
		\label{figG}
	\end{figure}
	The second derivative of eq.(\ref{eq20}) gives the collapsing acceleration as
	\begin{equation}
		\frac{\ddot{a}}{a}=\frac{\left(\frac{1}{\beta }-1\right) \beta }{9 \left[18^{\frac{\beta }{3-2 \beta }} \left(\frac{M}{r_{0}^3}\right)^{\frac{\beta }{3-2 \beta }}-\frac{t\beta  }{3}+\frac{  t_{0}\beta}{3}\right]^2}
		\label{eq26}
	\end{equation}
 \begin{figure}[h]
	\centering
	\includegraphics[scale=0.95]{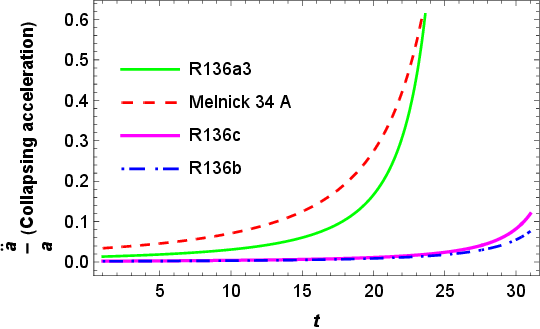}
	\caption{Variation of the collapsing acceleration $\frac{\ddot{a}}{a}$ versus time coordinate $t$ for the four massive stars .}
	\label{figH}
\end{figure}
	Substituting eq.(\ref{eq20}) into eq.(\ref{eq12}), the Kretschmann curvature assume the following form
\begin{equation}
	\mathcal{K} =\frac{12 \left(2-2 \beta +\beta ^2 \right)}{\left[2^{\frac{\beta }{3-2 \beta }} 27^{\frac{1}{3-2 \beta }} \left(\frac{M}{ r_{0}^3}\right)^{\frac{\beta }{3-2 \beta }}- t  \beta  + {t_0} \beta \right]^4}
	\label{eq27}
\end{equation}
\begin{figure}
	\begin{subfigure}[a]{0.4\textwidth}
		{\includegraphics[width=7cm]{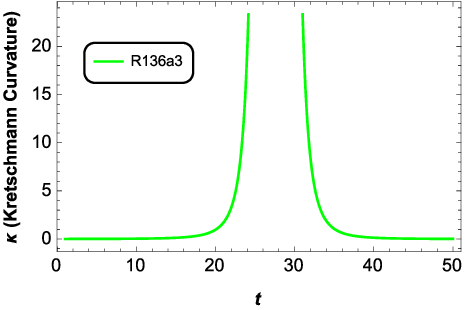}}
		\label{fig9}
	\end{subfigure}
	\begin{subfigure}[a]{0.4\textwidth}
		{\includegraphics[width=7.2cm]{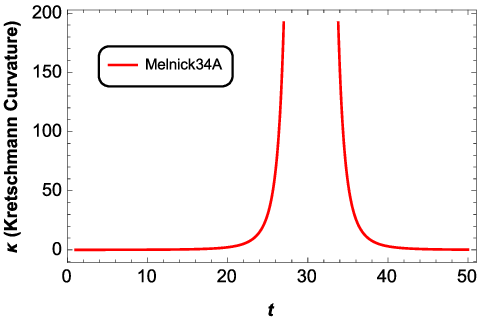}}
		\label{fig10}
	\end{subfigure}
	\begin{subfigure}[c]{0.4\textwidth}
		{\includegraphics[width=7cm]{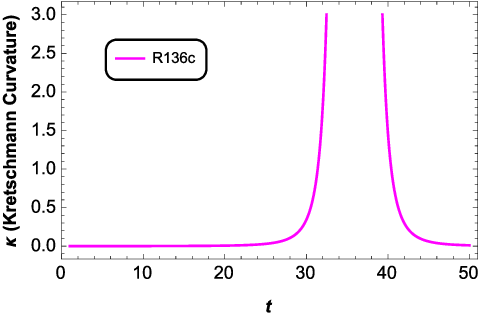}}
		\label{fig11}
	\end{subfigure}
	\begin{subfigure}[c]{0.4\textwidth}
		{\includegraphics[width=7cm]{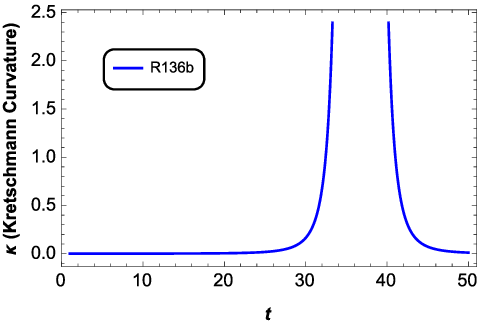}}
		\label{fig12}
	\end{subfigure}
	\caption{Variation of the Kretschmann scalar $(\mathcal K)$ versus time coordinate $t$ for the four massive stars .}
	\label{figI}
\end{figure}
	It can be seen from eqs.(\ref{eq20})-(\ref{eq27}) that all the quantities namely $a$, $\rho$, $p$, $m$, $\dot{m}$, $m'$,  $\frac{\ddot{a}}{a}$, and $\mathcal{K}$  are all explictly obtained in terms of stellar-mass $M$ and model parameter $\beta$.

	\subsection{The model parameter and comparison with relevant stellar-data}\label{subsec4.2}
	We consider the astrophysical data of four known massive stars namely $R136a3$, $Melnick$, $R136c$, and $R136b$ (with their known masses, radii, and present age), and compare the robustness of our model. These data point values are utilized to estimate the numerical value of model parameter $\beta$ for these stars (see Table- \ref{table1} ). By using astrophysical data and the numerical value of $\beta$, we have calculated the parameters $p, \rho$, $\Theta, \mathcal{K}, \frac{\ddot{a}}{a}$ for massive stars and present their trends graphically. In Table-\ref{table2}, we compare the present age of stars ($t_{age}$) with their collapse time $t_s$ (time of formation of BH discussed in section-\ref{sec5}) and theoretically predict their life-span. In Table-\ref{table2}, we also present the comparison between $t_s$ and $t_{AH}$ and discuss the nature of the singularity, where $t_{AH}$ is the time of formation of the apparent horizon discussed in section-\ref{sec5}. Using these data sets, we discuss the energy conditions (Null, Weak, Strong, and Dominant) and stability criteria for our model (see Table-\ref{table3}). 
	\begin{table}[h]
		\caption{Numerical value of model parameter $\beta$ for  the known astrophysical data of massive stars (based on their masses and radii) }
		\setlength{\tabcolsep}{3
			\tabcolsep}
		\centering
		\begin{tabular}{ *{4}{c} }
			\toprule
			\textbf{Massive Star} & \textbf{$Mass (M_\odot)$} & \textbf{$ Radius (R_\odot)$} & \textbf{$\beta$}\\
			\midrule
			R136a3\cite{star1}-\cite{star1c}  & 155 & $25.5$ &	0.27515	\\
			\addlinespace 	\addlinespace

			Melnick 34A \cite{star2}-\cite{star2a} & 148&  $19.3$ &0.16768 \\ 	\addlinespace 	\addlinespace
			
			R136c\cite{star3} & 142&  40.7 & 0.44148 \\ 	\addlinespace 	\addlinespace

			R136b \cite{star4}-\cite{star4a} &117&    40   &0.46275\\
			\bottomrule
		\end{tabular}
		\label{table1}
	\end{table}
 
	\section{Singularity analysis} \label{sec5}
	\subsection{Apparent horizon}\label{subsec5.1}
	Apparent horizons are the space-like surfaces with future-pointing converging null
	geodesics on both sides of the surface. Self-gravitating systems generally end with the formation of a space-time singularity as a consequence of gravitational collapse (GC), which is defined by the divergence of curvature $\mathcal{K}$ and the energy density $\rho$. The development of trapped surfaces in space-time as the collapse proceeds then characterizes the various consequences of GC in terms of either a BH or a naked singularity (NS). No portions of space-time are initially trapped when an object begins to collapse due to its own gravity, but once certain high densities are attained, trapped surfaces form, and a trapped region develops in space-time \cite{S1}-\cite{S3}. It is this part of space-time that evolves, eventually forming the BH singularity in a collapsing configuration, and before it settles into its final state, the boundary of the trapped surface is marked by the presence of an apparent horizon. The “apparent horizon” is the outer boundary of the "trapped surface" which is the union of all trapped surfaces \cite{a1}-\cite{S4}. The apparent horizon typically develops between the time of formation of space-time singularity and the time at which it meets the outer Schwarzschild event horizon, and the singularity can be either causally connected or disconnected from the outside universe, which is decided by the pattern of trapped surface formation as the collapse evolves \cite{S6}. 
	\par
	
	For the space-time metric (\ref{eq1}), the apparent horizon(AH) is characterized by \cite{S6}-\cite{S10}
	\begin{equation}
		R_{,i} R_{,j} g^{ij} = \left(r \dot{a}\right)^2 - 1 = 0
		\label{eq28}
	\end{equation}
	where the comma (,) denotes the  partial derivatives and 
	$R(t,r) \equiv r a(t)$ denotes the geometric radius of the 2-spheres.
	\par 
	Let us assume that initially at $(t_{0},r_{0})$ the star is not trapped, then
	\begin{equation}
		R_{,i} R_{,j} g^{ij}|_{(t_{0},r_{0})} = \left(r_{0} \dot{a}(t_{0})\right)^2 - 1 < 0
		\label{eq29}
	\end{equation}
	Further, consider that at time $t= t_{AH}$ the whole star collapses inside the apparent horizon $r=r_{AH}$, then it follows from  (\ref{eq28}) that 
	\begin{equation}
		r_{AH}^2 \dot{a}^{2}(t_{AH}) -1=0
		\label{eq30}
	\end{equation}
	where $(t_{AH}, r_{AH})$ are coordinate on the apparent horizon surface. Now, using eq.(\ref{eq20}) in eq.(\ref{eq30}), we obtain
	\begin{equation}
		\begin{split}
			r_{AH} =3 \left[18^{\frac{\beta }{3-2 \beta }} \left(\frac{M}{r_{0}^3}\right)^{\frac{\beta }{3-2 \beta }}+\frac{t_{0}\beta  }{3}-\frac{t_{AH}\beta  }{3}\right]^{\frac{1}{2} \left(2-\frac{2}{\beta }\right)}
			\label{eq31}
		\end{split}
	\end{equation}
	Solving eq.(\ref{eq31}) for  $t_{AH}$, we obtained
		\begin{equation}
		t_{AH} = \frac{1}{\beta}\Bigl[2^{\frac{\beta }{3-2 \beta }} 27^{\frac{1}{3-2 \beta }} \left(\frac{M}{r_{0}^{3}}\right)^{\frac{\beta }{3-2 \beta }}- 3^{\frac{1}{1-\beta }} \left(\frac{1}{r_{AH}^{2}}\right)^{\frac{\beta }{2-2 \beta }}+  t_{0} \beta \Bigr]
			\label{eq32}    
		\end{equation}
	 The eq.(\ref{eq32}) describes the time of formation of the apparent horizon surface. The approximate numerical value of $t_{AH}$ for massive stars are estimated in table-\ref{table2} (by choosing coordinate value $r_{AH}=1$).
	\subsection{Occurrence of black hole}\label{subsec5.2}
	From eq.(\ref{eq21}) and (\ref{eq27}) one can observe that the density($\rho$) and Kretschmann curvature($\mathcal{K}$) diverge at the $t=t_{s}$, where 
	\begin{equation}
		t_{s} = \frac{2^{\frac{\beta }{3-2 \beta }} 27^{\frac{1}{3-2 \beta }} \left(\frac{M}{r_{0}^3}\right)^{\frac{\beta }{3-2 \beta }}+\beta t_{0}}{\beta }
		\label{eq33}
	\end{equation}
	In order to examine the nature of the singularity of collapsing stars, we compare the collapse time $(t_s)$ and time formation of AH $(t_{AH})$ and see whether the singularity is covered by 	AH or not. Thus, from eq.(\ref{eq32}) and (\ref{eq33}), we obtained
	\begin{equation}
		t_{AH} - t_{s} = -3^{\frac{1}{1-\beta }} \left(\frac{1}{r_{AH}^2}\right)^{\frac{\beta }{2-2 \beta }}
		\label{eq34}
	\end{equation}
	In the BH scenario, the apparent horizon forms at a stage earlier than the singularity formation. The outside event horizon then entirely covers the final stages of collapse when the singularity forms, while the apparent horizon inside the matter evolves from the outer shell to reach the singularity at the instant of its formation \cite{S6},\cite{S10}. The eqs.(\ref{eq32})and  (\ref{eq33}) show that $t_{AH}$ and $t_s$  both give finite time durations. Further equ.(\ref{eq34}) and Table (\ref{table2}) show that the apparent horizon forms earlier than the singularity formations that is, $t_{AH} < t_s$, and the singularity is covered by the horizon surface. Thus, with the analysis of above arguments, we conclude that the collapse of massive stars would inevitably lead to black hole formation as their end-state.
 
	\captionsetup{labelsep=newline,
		singlelinecheck=false,
		skip=0.333\baselineskip}
	\newcolumntype{d}[1]{D{.}{.}{#1}} 
	\renewcommand{\ast}{{}^{\textstyle *}} 
	\renewcommand\arraystretch{1.5}
	\begin{sidewaystable}[h]
		
		\caption{The comparison between present age and life-span of the star, the formation of AH, and the nature of the singularity for known massive stars } 
		\label{table2}
		\begin{tabular}{p{2.9cm}p{3.8cm}p{4cm}p{4cm}p{4.5cm}}
			\hline\specialrule{0.9pt}{0pt}{0pt}
			
			\textbf{Massive Star}  & present age \newline (Myr)&   Star’s Life span $(t_{s})$\newline (Myr)  &  $t_{AH}$\newline (Myr)  & Nature of Singularity\newline ($t_{AH}<t_{s}$) \\
			
			\hline\specialrule{0.6pt}{0pt}{0pt}
			\addlinespace \addlinespace
			
			R136a3  & 1.28 &	27.5804 &  11.0354     & BH \\
			\addlinespace 	\addlinespace

			Melnick 34A  & 0.5 &30.3897& 8.0667  & BH\\ 	\addlinespace 	\addlinespace
			
			R136c & 1.8  &35.8992& 19.705   & BH\\ 	\addlinespace 	\addlinespace

			R136b & 1.7 &36.7381&  20.0368   & BH\\
			\addlinespace 
			\hline\specialrule{0.6pt}{0pt}{0pt} 
		\end{tabular}
	\end{sidewaystable}
	\subsection{ Life span of massive stars}\label{subsec5.3}
Stars are born of gaseous clouds in the interstellar medium and evolve throughout their lifetimes from early to final stages. They are made of massive spheres composed of the most abundant elements in the universe—H and He—held together by their own gravity.  When massive stars burn through their nuclear fuels (H and He), they undergo a series of fusion reactions in their cores. Eventually, they develop an iron core that cannot be further fused into heavier elements, and the core cannot support itself against gravitational collapse.  The life cycle of a star, from birth in the giant molecular clouds, through the active phase involving nuclear fusion as an energy source, and ultimate death, principally depends on its mass \cite{stellar1}. The final stages of low-mass and high-mass stars are also different: low-mass stars die forming a red giant and then a white dwarf, while high-mass stars explode as supernovae, leaving behind a neutron star or black hole.  Massive stars also have some specific features in their spectra and have characteristic life spans in the order of a few million years and this is still sufficient time for their stellar clouds to carry away a significant proportion of the total stellar mass \cite{stellar2}.
	
\par
The life span of a massive star depends primarily on its mass, and there is a clear correlation between a star's mass and its evolutionary path \cite{stellar3}-\cite{stellar4}. Massive stars, those with masses significantly greater than our Sun, follow a unique evolutionary path that ultimately leads to the formation of black holes when they exhaust their nuclear fuel. The most massive stars with mass $>M_\odot$ have a life span measured in millions of years and they have shorter lifespans as compared to less massive stars. One can see from eq.(\ref{eq33}) that the life-span ($t_s$) of massive stars is a function of their masses and it decreases with it (see Table-\ref{table2}). Thus, our model corresponds to the evolutionary stages of stellar systems and theoretically predicts the possible life-span of massive stars- \textit{R136b, R136c,	Melnick 34A, R136a3} in a table-\ref{table2}
	
	\section{Energy Conditions}\label{sec6}
	A physically reasonable solution of EFEs should satisfy certain energy conditions namely- Weak, Null, and Strong. Since the equation of state (EoS) in any stellar interior is still an unexplored arena, the energy conditions in gravitational theories are created to decode as much information as possible from classical general relativity without the administration of a specific EoS for the stress-energy \cite{e1}.  When the stress-energy tensor is under specific constraints, there is frequently a linear connection between $\rho$ and $p$. In this section, we describe the energy conditions to be applied so that the model	becomes physically acceptable.  
	
\subsection{Weak energy condition (WEC)}\label{sub6.2} 
\begin{figure}[h!]
	\centering
	\includegraphics[scale=0.64]{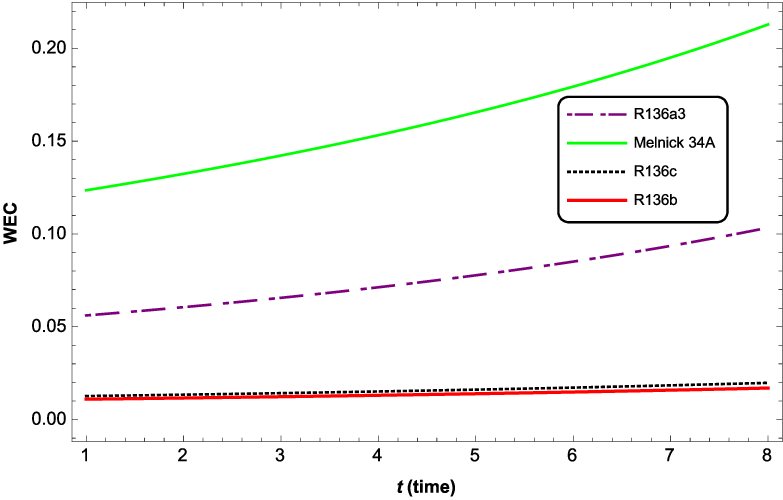}
	\caption{Nature of $WEC$ versus the time  $(t)$ given in table \ref{table3} for the four massive stars.}
	\label{fig13}
\end{figure}

The weak energy condition must be satisfied for the interior fluid distribution of the stellar system and obeyed for a physically acceptable model of gravitational collapse. According to WEC, the energy density measured by any time-like observer is non-negative ($\rho \geq 0$), and 
pressure cannot be so negative that it dominates the energy density ($\rho+p \geq 0$). From eqs. (\ref{eq21})-(\ref{eq22}), we see that our model satisfied the WEC and a graphical visualization of this condition is shown in figure (\ref{fig13}) for the model of four massive stars.

\subsection{Null energy condition (NEC)}\label{sub6.1} 
\begin{figure}[h!]
	\centering
	\includegraphics[scale=0.80]{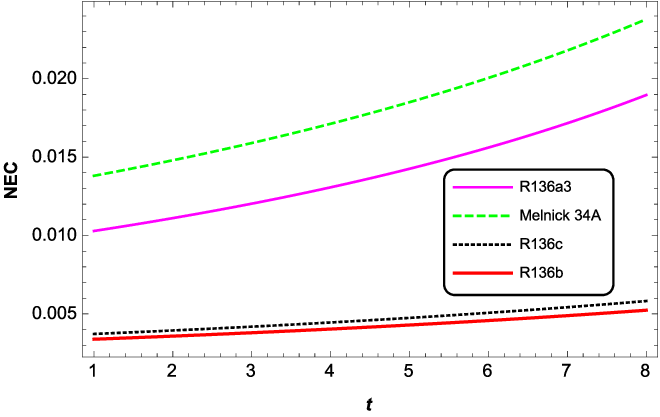}
	\caption{Nature of $NEC$ versus the time  $(t)$ given in table \ref{table3} for the four massive stars.}
	\label{fig14}
\end{figure}

The null energy condition must be obeyed for a physically acceptable model of gravitational collapse and according to this  $\rho+p \geq 0$. The NEC is satisfied for our model as shown in figure (\ref{fig14}).

\subsection{Strong energy condition (SEC)}\label{sub6.3} 
\begin{figure}[h!]
	\centering
	\includegraphics[scale=0.65]{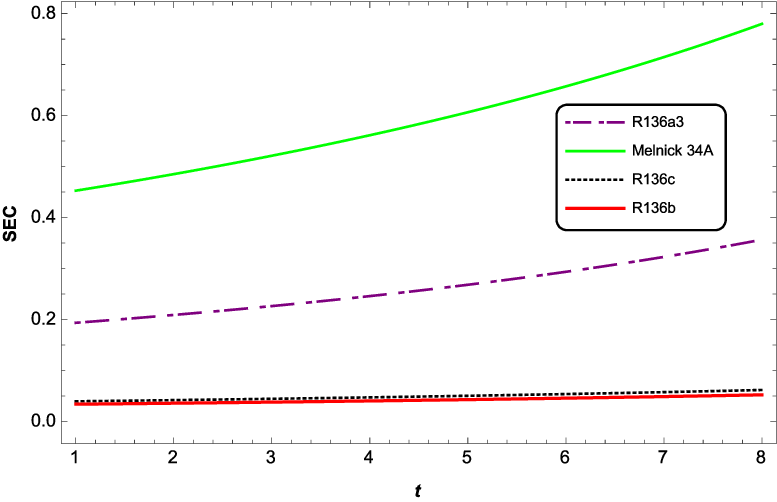}
	\caption{Nature of $SEC$ versus the time  $(t)$ given in table \ref{table3} for the four massive stars.}
	\label{fig15}
\end{figure}

According to SEC, $\rho+p \geq 0$ and $\rho+3 \lvert p \lvert >0$. With the critical analysis of eqs.(\ref{eq21})-(\ref{eq22}), we see that  $\rho+3 \lvert p \lvert >0$
that is, SEC is also satisfied in our model as can be seen in figure(\ref{fig15}).
	\section{Stability Analysis}\label{sec7}
	An approach for investigating potentially the stability analysis of the stellar model is the cracking method. For Self-gravitating stellar compact objects, the concept of cracking method for fluid distribution was first studied by Herrera \cite{h1}. This condition is used to determine the stability of a configuration of collapsing fluid. This condition suggests that for any stellar model to be physically acceptable, the speed of sound needs to satisfy causality conditions $0 < v_s^2 < 1$ that is, the speed of sound propagation inside the star must be less than the speed of light $c$ (taken $c = 1$), where $v_s$ is the speed of sound propagation inside the star and is obtained as
	\begin{equation}
		v^2_{s}= \lvert\frac{dp}{d\rho}\lvert=\lvert \frac{2\beta}{3}-1 \lvert
		\label{eqs37}
	\end{equation}
	We determined the numerical value of sound's speed($v_s$) for the four known massive stars (see table (\ref{table3})) and concluded that they satisfy the stability criteria for our model.
	\captionsetup{labelsep=newline,
		singlelinecheck=false,
		skip=0.333\baselineskip} 
	\newcolumntype{d}[1]{D{.}{.}{#1}} 
	\renewcommand{\ast}{{}^{\textstyle *}} 
	\renewcommand\arraystretch{1.5} 
	\begin{table}[h]
		\caption{The Energy Conditions and Stability factor $(v_{s})$ for known massive stars for model parameter given in Table \ref{table1}}
		\label{table3}
		
		\begin{tabular}{p{2cm}p{2.4cm}p{4cm}p{4cm}p{2cm}}
			\hline\specialrule{0.4pt}{0pt}{0pt}
			\textbf{Massive Star} &  $(\rho \geq0)$  & $(\rho+p \geq0)$ &  $(\rho+3\lvert p \lvert \geq 0)$ & Stability \newline $(v_{s}^2 <1)$ \\  
			\hline\specialrule{0.4pt}{0pt}{0pt}
			\addlinespace
			$R136a3$  & $\frac{39.6262}{(27.5804 - t)^2}$ &  $\frac{7.26876}{(27.5804 - t)^2}$ & $\frac{136.699}{(27.5804 - t)^2}$ & $0.816567$  \\	\addlinespace \addlinespace \addlinespace \addlinespace \addlinespace

			$Melnick34A$   & $\frac{106.699}{(30.3901 - t)^2}$& $\frac{11.9279}{(30.3901 -t)^2}$ &  $\frac{284.313}{(30.3901 - t)^2} + \frac{106.699}{(30.3901-t)^2}$   & $0.888213$ \\
			\addlinespace \addlinespace \addlinespace 
			
			$R136c$  & $\frac{15.3921}{(35.8988 - t)^2}$& $\frac{4.53022 t^2-325.259 t+ 5838.21}{(35.8988 - t)^4}$& $\frac{47.97779 t^2-3444.7 t + 61830.4}{(35.8988 - t)^4}$ & $0.70568$\\
			\addlinespace \addlinespace \addlinespace 
			
			$R136b$ & $\frac{14.0097}{(36.7375 - t)^2}$& $\frac{4.32199 t^2-317.558 t+ 5833.16}{(36.7375 - t)^4}$  & $\frac{43.0728 t^2-3164.78  t+58133.1}{(36.7375 - t)^4}$ & $0.6915$ \\
			
			\\		\hline\specialrule{0.4pt}{0pt}{0pt}
		\end{tabular}
	\end{table}
	
	\section{Discussion and Concluding Remarks}\label{sec9}
	
 The present study aims to discuss the singularity formation (BH) during the homogeneous gravitational collapsing phase of stellar systems and provide an appropriate model that incorporates known astrophysical stellar data.  The exact solutions and singularity formation studies are crucial in general relativity, and very few models provide physically interesting results in an astrophysical scenario. In this work, we have investigated the gravitational collapse of massive perfect fluid spheres within the framework of classical general relativity. The space-time of the interior matter distribution is described by the homogeneous and isotropic FLRW metric in which the particle trajectories are geodesics. Since $\Theta^{-1}$ is the only natural time scale in the model, we chose a power-law parametrization of the expansion scalar of the form $\Theta = -{a}^{-\beta}$. This parametrization allowed us to solve the governing field equation which subsequently determined the gravitational and dynamical evolution of the collapse process.  In order to assess the astrophysical relevance of our model in terms of graphical depiction, we have taken into consideration the massive stars namely, $R136a3$, $Melnick$, $R136c$, and $R136b$  and throughout the work, all graphs are drawn for these stellar candidates. We expected that the dynamics of the model would be sensitive to the model parameter, $\beta$ which is estimated numerically for known data of aforementioned stars. For this collapse scenario, the end-state will always be a black hole, i.e., the horizon forms in advance of the formation of the singularity. In addition, we were able to calculate the life-span of some well-known stellar candidates by utilizing their masses, present observed ages, and the time of formation of the singularities.  Our model is subjected to various physical tests based on regularity, causality, and stability and satisfies the energy conditions (Figures \ref{fig13}, \ref{fig14}, \ref{fig15}). 
	\par
	Some of the important graphical features of our model regarding the collapsing configuration are as follows-
	\par
	$\bullet$  The expansion scalar $\Theta$ in a collapsing system increases over time and tends to infinity at the center r = 0 (Fig-\ref{figA}). The negative sign of $\Theta$ represents the motion of collapsing fluids towards the center.
	\par
	$\bullet$ The scale factor $a$ and radius $R( = ra)$ are regular and finite inside the system and as expected are monotonically decreasing in nature (Fig-\ref{figB}).  It can be also observed from Eq.(\ref{eq20}) that the collapse attains central singularity ($a = 0$) at $t \rightarrow t_s$.
	\par
	$\bullet$ The energy density $\rho$, pressure $p$ and Kretschmann curvature $\mathcal{K}$ given in eqs.(\ref{eq21}),(\ref{eq22}) and (\ref{eq27}) are increasing in nature  and diverge at $t\rightarrow t_s$ as shown in figures \ref{figC}, \ref{figD}, and \ref{figI} . The negative sign of $p$ indicates the pressure towards the center ($r = 0$) during the collapsing configuration.  These figures explicitly show that the collapse of massive stars attains BH singularity in finite duration $t= t_s $ as their end state.
	
	\par
	$\bullet$ The Collapse acceleration $\frac{\ddot{a}}{a}$ increases over time which indicates the accelerating phase of collapsing configuration (Fig-\ref{figH}).  The collapsing acceleration is observed to be rapidly high at $t \rightarrow  t_s$ (see eq.(\ref{eq26})).
	 \par
	 $\bullet$ The profile of collapsing mass $m(t,r)$ shows that it is regular, positive, and decreasing with $t$ and $r$ (Fig-\ref{figE}) and becomes infinitesimally small ($ 10^{-34}$ order) at $t \rightarrow t_s$. The rate of change of mass, $\dot{m}$ and its gradient, $m’$ are decreased with $(t,r)$ which shows the loss of mass during the collapse process (Fig- \ref{figF} and \ref{figG}). In classical physics, an absolute ground state is defined by $E=0$, meaning if a star truly becomes BH, then $E= M c^2 = 0$.; i.e., $M\rightarrow 0$ (may be called as expected mass-singularity).  The mathematical "black hole" solution accurately predicts that a black hole (BH) may have an infinitesimally small mass\cite{mitra02}. In general relativity, a $M \rightarrow 0$ event doesn't imply the absence of matter, as gravitational mass comprises all energy sources, including negative self-gravitational energy.   Thus, such phenamenon may indicate extreme self-gravitation, offsetting other energy sources like protons, neutrons, and internal energies like heat and pressure\cite{mitra03}. 

	\par
	The present study found that our suggested model is highly significant for realistic stellar systems and that it may be employed to explain the existence of massive stars. We have provided a fully consistent general relativistic treatment to describe collapse scenarios of stellar objects with known masses and radii. It is also of interest to point out that our model describes physically realizable stellar structures without resorting to exotic matter distributions such as dark energy and dark matter.  Looking to future studies, it would be of interest to relax the conditions applicable to a perfect fluid and to incorporate anisotropic stresses, shear, density inhomogeneities and dissipation in the stellar interior.  Our approach in this work can also be applied to stellar modeling and the study of gravitational collapse in modified gravitational theories.
	\par
	\textbf{Acknowledgment:} The authors AJ, RK, and SKS are acknowledged by the Council of Science and Technology, UP, India vide letter no. CST/D-2289. In addition, the authors RK and SKJP are  thankful to IUCAA for its visiting associateship program which assists in many ways.

\end{document}